\documentclass[twocolumn,english,aps,pra,longbibliography,superscriptaddress, floatfix]{revtex4-2}
\usepackage[utf8]{inputenc}
\usepackage{xcolor}
\usepackage{amssymb}
\usepackage{amsmath}
\usepackage{braket}
\usepackage{txfonts,stmaryrd}
\usepackage{graphicx}
\usepackage{enumerate}
\usepackage{algorithm}
\usepackage{algpseudocode}
\usepackage{makecell}
\usepackage{soul}

\usepackage{amssymb}\usepackage[colorlinks=true,urlcolor=blue,citecolor=blue,linkcolor=blue]{hyperref}

\usepackage{comment}
\usepackage{natbib}

\begin{document}

\title{Genuine Multipartite Entanglement in Quantum Optimization}
\date{\today}

\author{Gopal Chandra Santra}
\email{gopalchandra.santra@unitn.it}
\affiliation{Pitaevskii BEC Center and Department of Physics, University  of  Trento,  Via Sommarive 14, I-38123 Trento, Italy}
\affiliation{INFN-TIFPA, Trento Institute for Fundamental Physics and Applications, Via Sommarive 14, I-38123 Trento, Italy}
\affiliation{Kirchhoff-Institut f\"ur Physik, Universit\"at Heidelberg, Im Neuenheimer Feld 227, 69120 Heidelberg, Germany}

\author{Sudipto Singha Roy}

\affiliation{Pitaevskii BEC Center and Department of Physics, University  of  Trento,  Via Sommarive 14, I-38123 Trento, Italy}
\affiliation{INFN-TIFPA, Trento Institute for Fundamental Physics and Applications, Via Sommarive 14, I-38123 Trento, Italy}
\affiliation{Department of Physics, Indian Institute of Technology (ISM) Dhanbad, IN-826004, Dhanbad, India}

\author{Daniel J. Egger}
\affiliation{IBM Quantum, IBM Research Europe – Zurich, S\"aumerstrasse 4, CH-8803 R\"uschlikon, Switzerland}

\author{Philipp Hauke}
\email{philipp.hauke@unitn.it}
\affiliation{Pitaevskii BEC Center and Department of Physics, University  of  Trento,  Via Sommarive 14, I-38123 Trento, Italy}
\affiliation{INFN-TIFPA, Trento Institute for Fundamental Physics and Applications, Via Sommarive 14, I-38123 Trento, Italy}

\begin{abstract}
The ability to generate bipartite entanglement in quantum computing technologies is widely regarded as pivotal.
However, the role of genuinely multipartite entanglement is much less understood than bipartite entanglement, particularly in the context of solving complicated optimization problems using quantum devices. It is thus crucial from both the algorithmic and hardware standpoints to understand whether multipartite entanglement contributes to achieving a good solution.
Here, we tackle this challenge by analyzing genuine multipartite entanglement---quantified by the generalized geometric measure---generated in Trotterized quantum annealing and the quantum approximate optimization algorithm. 
Using numerical benchmarks, we analyze its occurrence in the annealing schedule in detail. We observe a multipartite-entanglement barrier, 
and we explore how it correlates to the algorithm's success. We also prove how multipartite entanglement provides an upper bound to the overlap of the instantaneous state with an exact solution. Vice versa, the overlaps to the initial and final product states, which can be easily measured experimentally, offer upper bounds for the multipartite entanglement during the entire schedule.  
Our results help to shed light on how complex quantum correlations come to bear as a resource in quantum optimization.

\end{abstract}

\maketitle

\section{Introduction}

Quantum effects are expected to play a central role in quantum algorithms designed to address specific computational tasks \cite{jozsa1998quantum}.
However, understanding their precise contribution to the algorithm's performance is a subtle question. 
For instance, a key question is whether entanglement is necessary or sufficient in pure-state quantum computing~\cite{jozsa2003role, ding2007review, noah2001good}. 
Among the various applications of quantum computers, quantum combinatorial optimization stands out as particularly promising~\cite{Abbas2024}. 
Here, the algorithm is tasked to find an optimal solution $x^*$, or a near-optimal one, of a classical optimization problem $\min f(x)$~\cite{bauza2024scaling}. 
Interestingly, in quantum optimization approaches, such as quantum annealing (QA)~\cite{hauke2020perspectives, rajak2023quantum}, the quantum approximate optimization algorithm (QAOA)~\cite{Farhi2014}, and variational quantum algorithms (VQA)~\cite{cerezo2021variational}, the initial state is typically a product state, while the solution state should encode a classical problem and thus (barring superpositions of degenerate solutions) may not contain any entanglement. 
Having both the initial and solution state with little to no entanglement raises the question of how much entanglement is genuinely required to make a quantum-optimization algorithm successful~\cite{hauke2015probing}.

In previous years, most analyses focused exclusively on bipartite entanglement~\cite{orus2004universality,pati2009role, hauke2015probing, dupont2022calibrating, dupont2022entanglement}, where the entanglement between two non-overlapping partitions (typically of equal size) of the quantum state is computed. However, the behavior of bipartite entanglement can vary significantly depending on the chosen partitions~\cite{Singha_Roy_2020, Singha_Roy_2021}. Additionally, a state may be entangled across all possible bipartitions or may exhibit global or multipartite entanglement, making it impossible to describe the total entanglement through contributions from individual bipartitions~\cite{De_Chiara_2018}.
Since quantum optimization is, in essence, concerned with navigating a many-body system through a quantum spin-glass transition~\cite{binder1986spin,tanaka2017quantum}, one may expect entanglement shared between multiple parties to play an important role. 
In the context of pure-state quantum computing, it is known that merely increasing bipartite entanglement with system size is insufficient for an exponential speed-up~\cite{jozsa2003role}. 
In contrast, multipartite entanglement shared among many parties can be a valuable resource for quantum information processing, such as in quantum secret sharing~\cite{Hillery1999}, multi-party quantum teleportation~\cite{Chen_2006}, quantum key distribution~\cite{Epping_2017}, quantum metrology~\cite{T_th_2012}, and measurement-based quantum computation~\cite{Briegel_2009}. 
However, its precise role in ensuring successful performance remains unclear for quantum algorithms in general~\cite{Bru__2011} and quantum-optimization in particular~\cite{hauke2015probing,santra2024squeezing}.

Here, we analyze the role of genuine multipartite entanglement in quantum optimization using the generalized geometric measure (GGM) of entanglement~\cite{shimony1995degree,barnum2001monotones,wei2003geometric, sende2010pra,biswas2014pra}.
The GGM is an attractive measure for multipartite entanglement as it is experimentally accessible~\cite{badzia2008experimentally}. 
From detailed numerical benchmarks, we find---similar to the bipartite entanglement barrier~\cite{hauke2015probing,dupont2022entanglement}---a multipartite entanglement barrier in quantum annealing. 
Initially, the circuit builds up multipartite entanglement as it approaches the minimal gap, where the system enters approximately a GHZ state between the two lowest-energy states. 
Subsequently, in successful optimization schedules, this multipartite entanglement is removed as the system approaches the solution state.  
We further analytically derive how the GGM in the optimization schedule is upper-bounded by the distances between the instantaneous eigenstate and both the initial and final product states, which can also be used as a proxy for bounding multipartite entanglement in experiments.
Through our work, we shed light on the role and existence of multipartite entanglement in quantum optimization, 
and its relation to the success probability of the algorithm.

Our studies are complementary to a vigorous ongoing effort to understand the role of quantum resources in quantum optimization. E.g., there are works that study whether quantum optimization still performs well with a cap on entanglement~\cite {meyer2000sophisticated, biham2004quantum,kenigsberg2006quantum,nest2013universal,dupont2022calibrating,sreedhar2022quantum,naseri2022entanglement} or if the success of the algorithm requires other quantum resources~\cite{chitambar2019quantum}.

The rest of this paper is structured as follows. In Sec.~\ref{sec: background}, we provide a general overview of how entanglement manifests in quantum optimization, along with a definition of the GGM. From Sec.~\ref{sec: maxcut using quantum annealing} onwards, we introduce quantum annealing and the MaxCut problem used in our simulation. Section~\ref{sec: fixed problem} details the results from Trotterized quantum annealing, focusing on success probability, bipartite entanglement, and the GGM. In Sec.~\ref{sec: entanglement barrier}, we explore the emergence of the entanglement barrier and its connection to success probability. Finally, we conclude in Sec.~\ref{sec: conclusion}.

\section{Background}\label{sec: background}

First, we briefly summarize earlier works that relate bipartite entanglement and the performance of certain classes of quantum algorithms. 
Thereafter, we motivate the necessity of analyzing multipartite entanglement properties in quantum algorithms and define a computable measure for it, the GGM of multipartite entanglement. 
Finally, we describe the quantum optimization algorithm and the problem we investigate.  

\subsection{Entanglement in Quantum Optimization}\label{sec: entanglement in quantum algo}

Various efforts have already demonstrated how bipartite entanglement manifests in quantum optimization algorithms. 
The presence of von Neumann entanglement entropy has been observed numerically during quantum annealing sweeps in the exact cover problem~\cite{orus2004universality}.
Furthermore, entanglement witnesses in experiments have detected non-zero entanglement during quantum annealing~\cite{lanting2014entanglement}.
Reference~\cite{hauke2015probing} explores a potential connection between bipartite entanglement and the success probability of adiabatic quantum optimization. 
It shows that high entanglement during the optimization does not necessarily mean a high success probability. 
On the contrary, in clean systems, significant final entanglement after slow sweeps suggests a superposition state rather than a separable ground state and thus implies a reduced probability of successfully finding the classical ground state.
Therefore, measuring bipartite entanglement alone is insufficient to gauge the efficiency of adiabatic quantum optimization.

The subtle role of bipartite entanglement is also observed in the variational quantum eigensolver (VQE), where distributing the entangling gates according to the problem's topology increases success rates and reduces runtime~\cite{valle2021quantum}. 
Moreover, the ADAPT variant of QAOA uses a non-standard mixer, which allows a larger amount of entanglement in the earlier part of the circuit and accelerates convergence in the later stages~\cite{chen2022much}. 
To further understand the role of bipartite entanglement, one can simulate a quantum algorithm with matrix-product states with finite bond dimension ($\chi$), since $\chi$ bounds the maximum bipartite entanglement.
Using this approach, reference~\cite{dupont2022calibrating} shows that the state fidelity in QAOA can be limited by $\chi$.
In particular, QAOA needs a bond dimension that scales exponentially with the system size $N$ to create high-fidelity states~\cite{dupont2022entanglement}.
However, increasing the bond dimension beyond a certain threshold does not appear to improve the fidelity~\cite{sreedhar2022quantum}. 
Also, reference~\cite{santra2024squeezing} demonstrates numerically, as well as on IBM Quantum hardware, how QAOA can generate high entangled states with multipartite entanglement witnessed through squeezing inequalities and the quantum Fisher information.

\subsection{Generalized Geometric Measure of Entanglement}\label{sec: GGM defintion}
{\color{blue}}
\begin{figure}
    \centering
    \includegraphics[width=\columnwidth,clip, trim= 200 90 250 50]{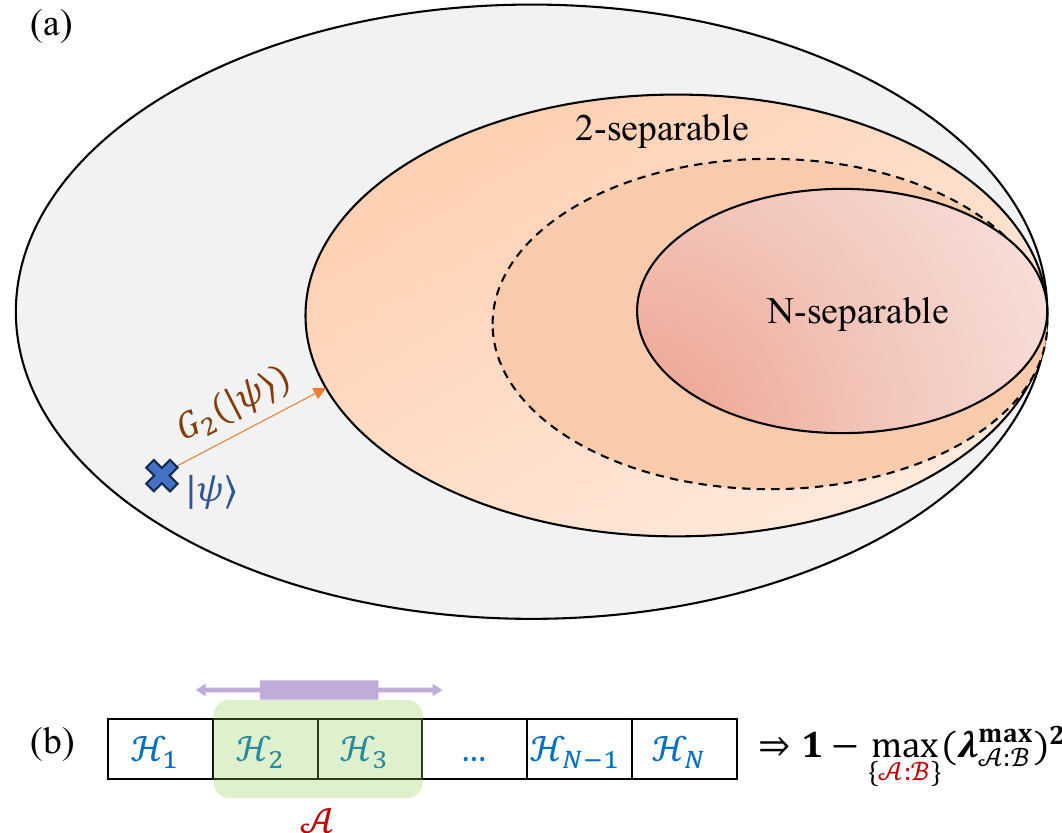}
    \caption{(a) Pictorial representation of the GGM: The GGM of a state is the minimum distance from all separable states, which coincides with the minimum distance from the 2-partite product states. (b) For pure states, instead of needing to minimize the distance of a given pure state to all 2-separable states, a much more efficient way exists to compute the GGM. Namely, it can be obtained from the maximum Schmidt coefficient across all possible bipartitions of the system. Sketched is a system described as a tensor product of local Hilbert spaces $\mathcal{H}_j$, bi-partitioned into sets $\mathcal{A}$ and $\mathcal{B}$.}
    \label{fig: pictorial GGM}
\end{figure} 
The GGM of entanglement generalizes the notion of geometric measure of entanglement and quantifies genuine multipartite entanglement~\cite{shimony1995degree,barnum2001monotones,wei2003geometric, sende2010pra,biswas2014pra}. 
Intuitively, the GGM measures the shortest distance between a quantum state and the set of product states [see Fig.~\ref{fig: pictorial GGM}(a)].
Formally, we can define a hierarchy of geometric measures $ G_k$ with $2\leq k \leq N$ of a $N$-party pure quantum state $\ket{\psi}$ as the minimum distance between $\ket{\psi}$ and the set $\mathcal{S}_k$ of $k$-party product states $|\pi\rangle=|\psi_1\rangle \otimes |\psi_2\rangle \otimes \dots \otimes |\psi_k\rangle$, given by
\begin{equation}
     G_k\left(|\psi\rangle\right)=1-\max_{|\pi\rangle \in \mathcal{S}_k}|{\langle \pi|\psi\rangle|}^2 .
\end{equation}
This distance measure resembles the Fubini--Study~\cite{bengtsson2017geometry} and the Bures metrics~\cite{Bures_1969}. 
The original geometric measure of entanglement $G_{k=N}$ measures the distance between $\ket{\psi}$ and fully separable states. 
At the other extreme, \(G_{k=2}\) is zero if $\ket{\psi}$ can be written as a product state across any  {\it single} bipartition of the Hilbert space.
Even if that is the case, bipartite entanglement can still be found in other bipartitions.
Since \(\mathcal{S}_k \subset \mathcal{S}_2\) for all \(k\), the GGM measures the distance between \(\ket{\psi}\) and the union of all forms of \(k\)-separable states ranging from fully separable to biseparable states.
The GGM thus captures the amount of genuine multipartite entanglement present in the system. 

One may think that the inclusion of all kinds of biseparable states makes the measure hard to compute.
However, one can show that for pure states the GGM takes the simple form [see Fig.~\ref{fig: pictorial GGM}(b)] 
\begin{equation}\label{eq: GGM_using_schmidt}
     G_2(\ket{\psi}) = 1 - \max_{\ket{\pi}_{\mathcal{A}:\mathcal{B}}\in \mathcal{S}_2 } |{_{\mathcal{A}:\mathcal{B}}}\langle \pi|\psi\rangle|^2=1 - \max_{\mathcal{A}:\mathcal{B}}({\lambda^{\max}_{\mathcal{A}:\mathcal{B}}})^2.
\end{equation} 
Here, $\ket{\psi}$ belongs to the Hilbert space $\mathcal{H}_{A_1}\otimes \mathcal{H}_{A_2} \otimes \dots \otimes \mathcal{H}_{A_N}$. 
The bipartition $\mathcal{A}:\mathcal{B}$ satisfies $\mathcal{A} \cup \mathcal{B}=\{1,2,\cdots, N\}$, and $\mathcal{A} \cap \mathcal{B}=\emptyset$~\cite{sende2010pra,biswas2014pra}. 
In other words, maximizing $|{_{\mathcal{\mathcal{A}:\mathcal{B}}}\langle} \pi|\psi\rangle|$ across the biparition $\mathcal{A}:\mathcal{B}$ is equivalent to finding the maximum  Schmidt coefficient, $\lambda^{\max}_{\mathcal{A}:\mathcal{B}}$, for that bipartition.
Finally, to compute the GGM, one has to square the global maximum of the Schmidt coefficient over all bipartitions $\mathcal{A}:\mathcal{B}$ of the system~\footnote{In the absence of simplifications such as symmetries, we need to calculate over the following number of bipartitions--- ${N \choose 1}+\dots+{N \choose \frac{N-1}{2}}=2^{N-1}-1$ for odd $N$ and ${N \choose 1}+\dots+{N \choose \frac{N}{2}-1}+{N \choose \frac{N}{2}}=2^{N-1}-1 + \frac{1}{2} {N \choose \frac{N}{2}} $ for even $N$.}.
Although, in general, the number of partitions grows exponentially with the system size, one can often use symmetry or bounds to estimate the GGM effectively.
Moreover, in the context of analyzing the ground state of local many-body Hamiltonians, it is often observed that the main contribution to GGM comes from the bipartitions of the system into contiguous blocks. Hence, focusing on a smaller number of blocks would be sufficient to obtain the exact value of GGM in these cases, and one can expect a similar approach to work in some cases of quantum optimization as well.
One can show that the theoretical maximum value of GGM is obtained when, for a given bipartition, all the Schmidt coefficients become equal $\frac{1}{\sqrt{d}}$, and thus we have $G_2^{\max}(\ket{\psi})=1-1/d$, where $d$ is the local Hilbert space dimension~\cite{footnoteMax}. Extending the GGM to mixed states is non-trivial, though it can still be calculated by exploiting symmetries~\cite{das2016generalized,wei2003geometric,vollbrecht2001entanglement}.

Reference~\cite{rossi2013scale} investigates the GGM in Grover's algorithm.
Here, the GGM initially increases and reaches a maximum in the first half of the optimal number of iterations and then decreases towards the end of the algorithm.
In the Bernstein--Vazirani algorithm, the presence of geometric entanglement in the initial state prevents it from reaching an optimal performance (the key resource for the Bernstein--Vazirani algorithm is coherence rather than entanglement, and maximally entangled states cannot be maximally coherent)~\cite{naseri2022entanglement}. One can thus expect that the presence of an exaggerated amount of multipartite entanglement can have a detrimental effect, and understanding the nature of such entanglement may help us design more efficient quantum algorithms.

\subsection{Quantum Annealing and its Trotterization}\label{sec: maxcut using quantum annealing}

Quantum annealing leverages the adiabatic theorem to solve combinatorial optimization problems.
In this algorithm, one initializes the quantum state in an easy-to-prepare ground state of a mixing Hamiltonian $H_M$. 
Subsequently, the instantaneous Hamiltonian
\begin{equation}\label{eq:quantum_annealing}
    H(t)=A(t) H_{\mathrm{C}} +B(t) H_\mathrm{M}
\end{equation}
is slowly evolved from $H_\mathrm{M}$ at $t=0$ to the cost Hamiltonian $H_\mathrm{C}$ at $t=T$, whose ground state encodes the solution to the target problem.
I.e., $B(t)$ is slowly reduced from $B(0) = 1$ to $B(T) = 0$, and $A(t)$ is slowly increased from $A(0) = 0$ to $A(T) = 1$ (for simplicity, we choose $A(t)=t/T$ and $B(t)=(1-t/T)$ throughout).
According to the adiabatic theorem, if the time evolution is slow enough, the quantum state always remains in the ground state of the instantaneous Hamiltonian and reaches the solution state at the end of the sweep~\cite{amin2009consistency, hauke2020perspectives}. 
For quantum effects to play a role, $H_\mathrm{M}$ is chosen to be non-commutative with $H_\mathrm{C}$, usually as a sum of Pauli-X operators $H_\mathrm{M}=-\sum_i X_i$ (if the cost Hamiltonian is written in the computational basis). 

In practice, quantum annealing requires a very long sweep to successfully reach the solution state, i.e., the corresponding circuits are deep and error-prone. 
To execute quantum annealing on noisy digital devices with limited coherence time, one can discretize the continuous evolution as a sequence of gates, resulting in a Trotterized quantum annealing (TQA). 
This approach is very similar to QAOA, which additionally optimizes the time steps in a closed loop between the quantum computer and a classical co-processor.
A first-order Trotter--Suzuki decomposition transforms the continuous sweep, up to $\mathcal{O}(\Delta t^2 )$,  into 
\begin{equation}
    e^{-i\Delta t H(t)} \simeq e^{-i\beta H_\mathrm{M}} e^{-i\gamma H_\mathrm{C}}\,,
    \label{eq:Hstep}
\end{equation}
where $\beta=(1-t/T)\Delta t$, and $\gamma=(t/T) \Delta t$. Discretizing the entire quantum annealing schedule into $p$ time steps requires $\Delta t=T/p$. 
The resulting discretized annealing schedule is 
\begin{equation} \label{eq: delta t}
    \gamma_s=\frac{s}{p} \Delta t, \quad \beta_s=(1-\frac{s}{p})\Delta t\,,
\end{equation}
where the time-factor of $t/T$ is replaced by the layer-number $s/p$. 
Crucially, unlike in quantum annealing, here $\Delta t$ and the total time $T$ can be modified independently by changing $p$, which hands us an additional control knob for Trotterized annealing.

\subsection{Maximum Cut problem} 

As a paradigmatic example, we focus on the Maximum Cut problem (MaxCut). 
MaxCut aims at partitioning the set of $N$ nodes $V$ in a graph $\mathcal{G}(V,E)$, such that the sum of the weights $\omega_{i,j}$ of the edges $(i,j)\in E$ traversed by the cut is maximum. 
By introducing Ising variables $z_i\in\{-1,1\}$ to describe which side of the cut node $i$ falls on, this task is mathematically equivalent to maximizing the cost function 
\begin{align}\label{eq:max_cut }
    \max_{z\in\{-1,1\}^N}\frac{1}{2}\sum_{(i,j)\in E}\omega_{i,j}(1-z_iz_j).
\end{align}
Since $z_i$ can also be seen as the eigenvalues of the Pauli $Z_i$ operator acting on a qubit $i$, the maximization is equivalent to finding the ground state of the Ising model~\cite{Lucas2014,albash2018adiabatic}
\begin{equation}\label{eq: Hamiltonian without symmetry breaking}
    \tilde{H}_\mathrm{C}=\frac{1}{\sqrt{N}} \sum_{i<j} J_{ij}Z_iZ_j.
\end{equation}
Here, $J_{ij}$ encode the edge weights of $\mathcal{G}$ with $J_{ij}=0$ if $(i,j)\notin E$. 

Due to the $\mathbb{Z}_2$ symmetry of $\tilde{H}_\mathrm{C}$, there are two equivalent degenerate solutions. 
A ground-state search may thus result in varying superpositions of the two solutions, whose entanglement content can depend significantly on the details of the annealing sweep.
To ensure that the entanglement we observe is solely a signature of the performance of the quantum algorithm, we remove this degeneracy by adding a small perturbation $fZ_0$ to $\tilde{H}_\mathrm{C}$ to break the symmetry. 
We choose $f=0.05$ for all problems (see App.~\ref{app: breaking degeneracy}).
The modified Hamiltonian is thus
\begin{equation}\label{eq: max_cut with symmetry breaking}
    H_\mathrm{C}=fZ_0+\frac{1}{\sqrt{N}} \sum_{i<j} J_{ij}Z_iZ_j\,,
\end{equation}
which will be the problem on which we study genuine multipartite entanglement.

\section{Genuine multipartite entanglement for a fixed instance of MaxCut}\label{sec: fixed problem}
\begin{figure}[hbt!]
    \centering
    \includegraphics[width=\columnwidth,clip, trim= 0 0 0 0]{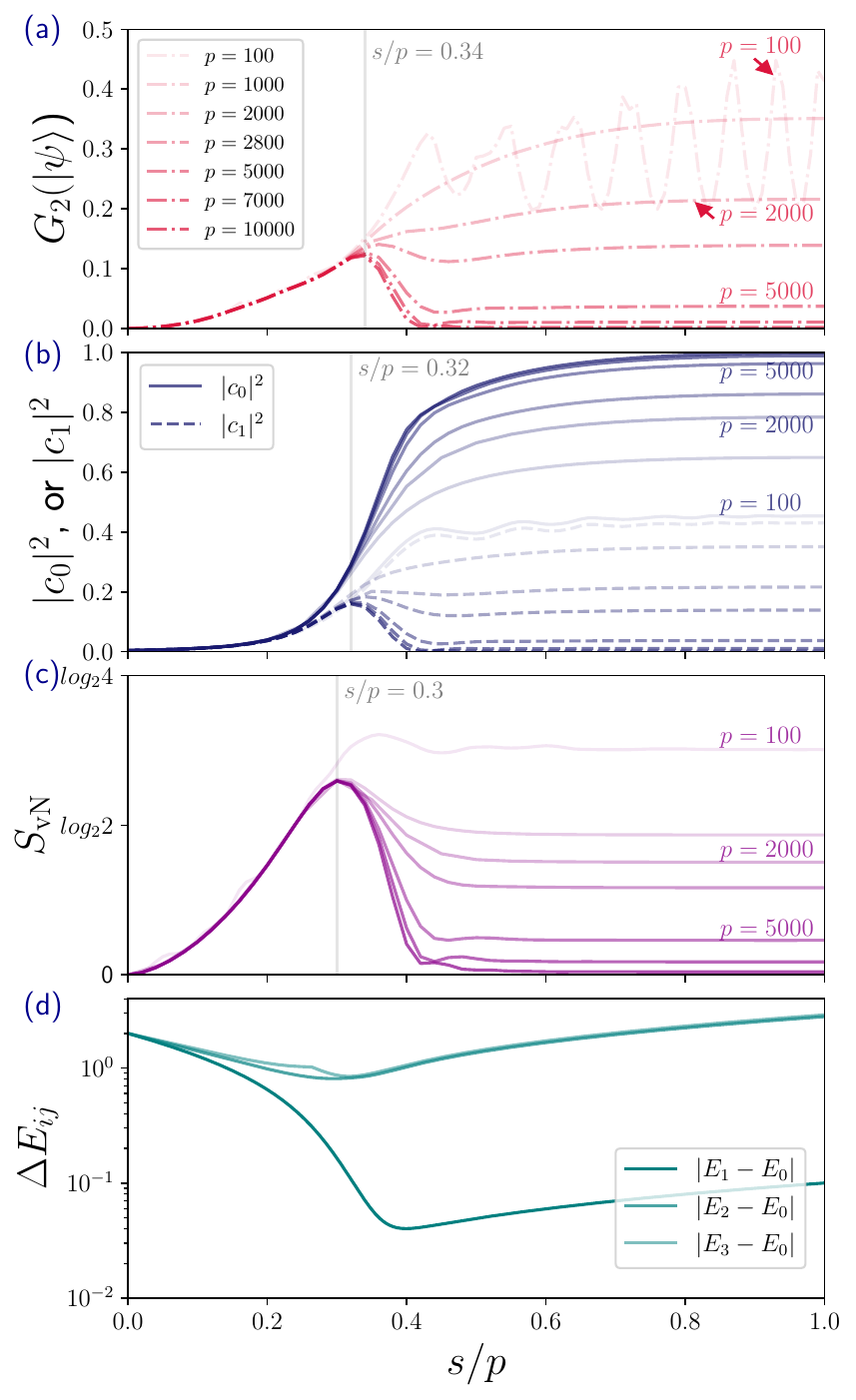}
    \caption{Results as a function of normalized layer number $s/p$ for a single MaxCut instance, obtained with different total numbers of TQA layers $p$. (a) At low values of $p$, the GGM saturates at high $s/p$, whereas for high $p$, a peak is observed around $s/p=0.34$, followed by a decrease and saturation at lower values. At fast sweeps ($p=100$), non-adiabaticity leads to superpositions involving excited states, the dynamical phases between which generate characteristic oscillation.
    (b) The overlap of the instantaneous state $\ket{\psi(s)}$ with the ground and first excited states of $H_\mathrm{C}$. 
    For slow sweeps, $|c_1|^2$ peaks around the point of maximum GGM.  
    (c) The von Neumann entanglement entropy measured in the middle of the graph 
    displays a peak for all $p$ considered but saturates to a lower value with increasing $p$. (d) Energy gaps between the instantaneous ground and higher excited states. The main features in the other panels appear in the vicinity of (but not exactly at) the closing of the low-lying gaps.}
    \label{fig: fixed_instance}
\end{figure}

In this section, we illustrate how, for an instance of the MaxCut problem, success probability, the von Neumann entanglement entropy, and the generalized geometric measure of multipartite entanglement behave during an optimization schedule. 
Unless stated otherwise, we choose the MaxCut problem on six-regular graphs on eight vertices, with edge weights $J_{ij}$ randomly chosen from a Gaussian distribution with mean $\mu=6$ and standard deviation $\sigma=3$.
For illustration purposes, in this section, we choose a fixed problem instance (details are in App.~\ref{app: problem instance}), and study the average behavior over many instances in the next section. 
We solve the problem with Trotterized quantum annealing and an increasing number of Trotter steps, or circuit layers, ranging from $p=100$ to $p=10000$. 
We select $\Delta t=0.19$, which allows us to achieve a good minimum energy across all layer counts, see App.~\ref{app: best dt}.
The total time is $p \Delta t$, which amounts to, e.g., $T=950$ for $p=5000$.

\subsection{Success probability for different sweep speeds}

Since more Trotter steps result in a more adiabatic evolution, we expect the success probability to increase with $p$.
We define the success probability as the degree of overlap between the final state $\ket{\psi(p)}_\text{TQA}$ and the true ground state $\ket{e_0}$ of $H_\mathrm{C}$. 
To understand the evolution of the state at each Trotter step, we evaluate the instantaneous probability of finding the ground state $\ket{e_0}$ and first excited state $\ket{e_1}$ of the final cost Hamiltonian at each layer $s$. 
Expressing the instantaneous state in the eigenbasis of $H_\mathrm{C}$,
\begin{equation}\label{eq: expansion in eigenstate}
    |\psi(s)\rangle_\text{TQA}=\sum_i c_i(s)\ket{e_i}\,,
\end{equation}
the instantaneous probability of being in the ground and first excited states are $|c_0(s)|^2$ and $|c_1(s)|^2$, respectively.
Here, the basis states $\{\ket{e_i}\}$ are ordered by increasing energy. 
The probability of reaching the final ground state, $|c_0(s=p)|^2$, defines the success probability of the algorithm. 

For all $p \in [1000,10000]$, the instantaneous ground-state probability $|c_0|^2$ increases with layer index $s$ and reaches the highest value at the end of the sweep $s=p$, see solid lines in Fig.~\ref{fig: fixed_instance}(b). 
As the number of layers $p$ increases, so does the success probability, i.e., slower sweeps are more successful. 
The first excited state probability $|c_1|^2$ (dashed lines) increases until $s/p\approx 0.32$, after which it decreases if $p$ is sufficiently large. 
This results in a peak, whose location is around the minimal energy gap between the first excited and the ground state of the instantaneous Hamiltonian [see Fig.~\ref{fig: fixed_instance}(d)].
The peak becomes sharper as the sweep speed is reduced, i.e., as $p$ increases.

For the smallest considered number of layers ($p=100$, i.e., $T=19$), an oscillation appears after crossing the minimum energy gap between the instantaneous ground and the first excited states. 
In such fast sweeps, the adiabatic conditions break down, causing the instantaneous state to become a superposition of multiple eigenstates. The dynamical phases between these lead to oscillations. One can see these also as the formation of defects, as described by the Kibble-Zurek mechanism (KZM)
~\cite{soriani2022three, miessen2024benchmarking,ivakhnenko2023nonadiabatic}.
As the adiabatic condition becomes valid for slower sweep speeds, the defects disappear, and we no longer see the oscillations.

\subsection{von Neumann entanglement entropy for bipartite entanglement}

Our modified MaxCut problem, given by Eq.~\eqref{eq: max_cut with symmetry breaking}, has by design a non-degenerate solution state. 
Therefore, we expect the ideal quantum anneal to reach this desired product state solution with vanishing entanglement. 
However, any realistic sweep within finite time has some degree of non-adiabaticity, thereby populating higher energy states, which generates undesired final-state entanglement~\cite{hauke2015probing}.
For comparison purposes to the GGM, we quantify the bipartite entanglement across the half--half bipartition of the system, i.e., $(0,1,2,3):(4,5,6,7)$, 
using the von Neumann entanglement entropy $S_{vN}=-\rm{Tr}\left[\rho_{\mathcal{A}} \log_2 (\rho_{\mathcal{A}})\right]$. 
Note that $S_{vN}$ depends on the chosen bipartition, whereas the GGM is optimized overall bipartitions.

As shown in Fig.~\ref{fig: fixed_instance}(c), we observe a maximum in $S_{vN}$ around the gap closing, reaching almost half of its theoretical maximum of $4\log_2(2)$ for the given subsystem size at $s/p\simeq 0.3$. 
Similar to reference~\cite{hauke2015probing}, slow sweeps (large $p$) reduce the entropy after this point as the quantum fluctuations produced by $H_M$ become weaker, while fast sweeps (small $p$) are unable to remove the entropy, resulting in a loss of success probability.

\subsection{Generalized geometric measure for multipartite entanglement}\label{sec: GGM in fixed problem}

For the fastest sweep considered, $p=100$, the GGM oscillates in accord with the oscillations of the overlaps with the ground and excited states, see Fig.~\ref{fig: fixed_instance}(a,b). The GGM peak near $s/p=0.34$ becomes more visible with slower sweeps, which resembles the observed peak in $|c_1|^2$.  
As the GGM follows qualitatively the same structure as the von Neumann entropy on the fixed bipartition, one may wonder if the optimal bipartition for the GGM is static or if the half--half bipartition captures the GGM on average. 
For the particular problem instance we considered, we found that the optimal bipartition remains constant for $0.3 \leq s/p \leq 0.8$ and is $\mathcal{A}=(2), {\mathcal{B}}=(0,1,3,4,5,6,7)$, which differs from the half--half bipartition used to compute $S_{vN}$, see App.~\ref{app: which bipartition}.  

\begin{figure}[hbt!]
    \centering
    \includegraphics[width=1\columnwidth, clip, trim= 0 0 215 35]{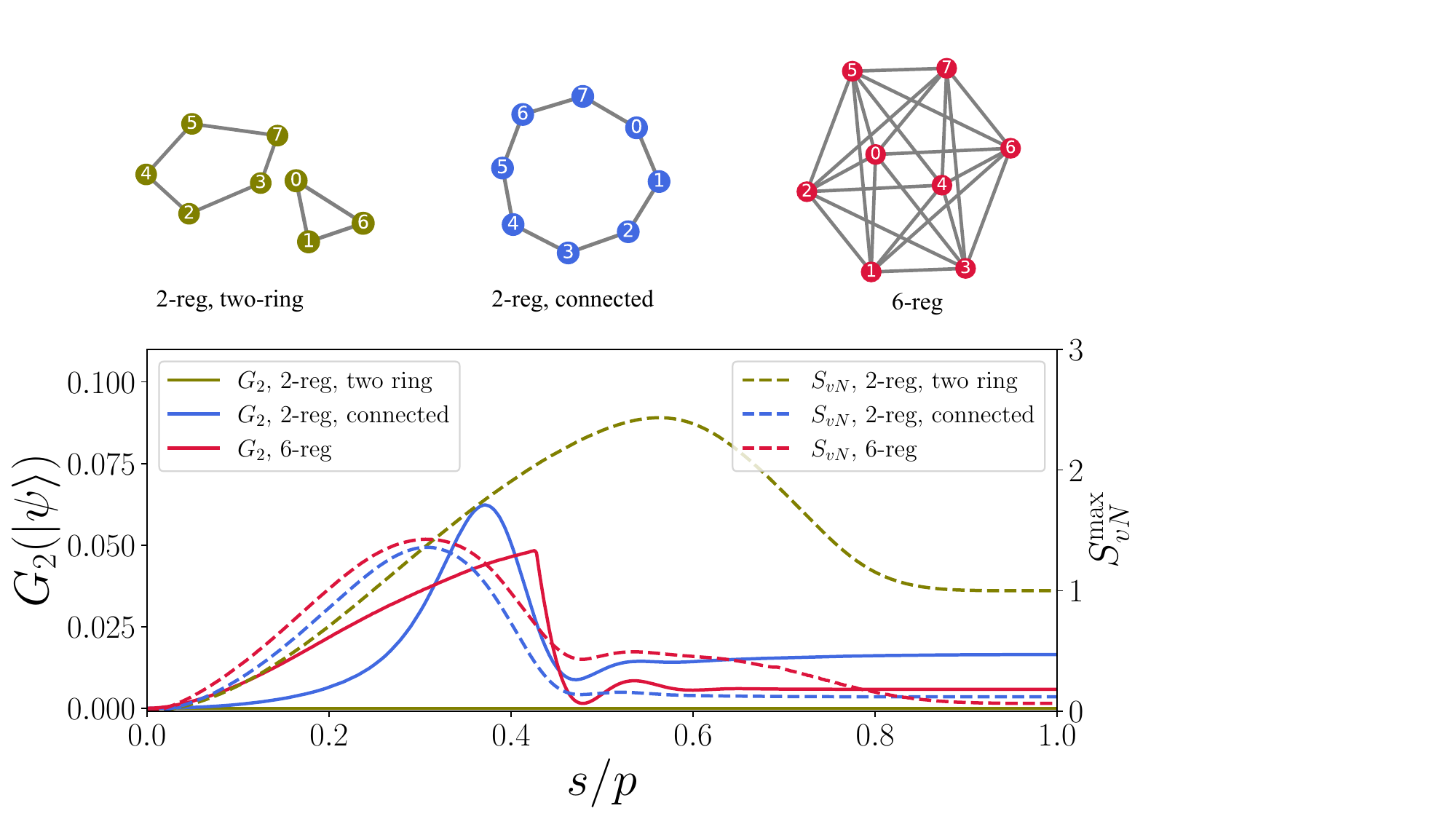}
    \caption{The von Neumann entanglement entropy of a half--half bipartition and the GGM can peak at different times in the annealing sweep. A two-regular graph can also have similarly strong entanglement as a six-regular graph. In a disconnected graph, the bipartite entanglement in a fixed bipartition can be large, yet multipartite entanglement can be zero. }
    \label{fig: GGM vs VNE}
\end{figure}

One might naturally question whether a graph's connectivity affects the value of the GGM, for instance, if a more connected graph would display greater entanglement. 
Perhaps surprisingly, even a two-regular graph, with just two edges per node, can exhibit nearly the same amount of GGM as a six-regular graph, see Fig.~\ref{fig: GGM vs VNE}. 
The same is true for the von Neumann entanglement entropy. 
Thus, it is crucial to understand in what sense bipartite and multipartite entanglement differ.

In general, bipartite entanglement for fixed bipartitions does not need to follow the behavior of the GGM, as we now show through some counter-examples.
As discussed in Sec.~\ref{sec: GGM defintion}, multipartite entanglement exists only when there is a non-zero bipartite entanglement for \textit{all} possible bipartitions. 
This implies that a state can have a large bipartite entanglement for a particular bipartition and yet can have a significantly smaller GGM, even zero. 
To exemplify the strong dependence of the von Neumann entropy on the bipartition, we consider $S_{vN}^{\rm max}$ as $S_{vN}$ maximized over $N \choose \lfloor N/2 \rfloor$ equal-sized (half--half) bipartitions, 
\begin{equation}
    S_{vN}^{\rm max}= \max_{|\mathcal{A}|=N/2: |{\mathcal{B}}|=N/2} S_{vN}[\mathcal{A}]\,.
\end{equation}

For the 6-regular graph studied above, we observe that the GGM peaks later than $S_{vN}^{\rm max}$, and the same happens in the example of a connected two-regular graph, see Fig.~\ref{fig: GGM vs VNE}.
This implies that in both cases, the bipartition corresponding to the GGM is not the same as the maximal bipartition of $ S_{vN}^{\rm max}$. 
An even stronger difference between half--half von Neumann entanglement entropy and GGM can be found in the example of a disconnected two-regular graph, see the ``2-reg, two ring data'' in Fig.~\ref{fig: GGM vs VNE}. In this extreme example, while the maximum bipartite entanglement entropy is large throughout the entire annealing schedule, the optimal bipartition of the GGM corresponds to the disconnected subgraphs, resulting in an identically vanishing GGM.

\subsection{Relation between state occupation probability and GGM }\label{sec: excited state and GGM}

\begin{figure}
    \centering
    \includegraphics[width=1\columnwidth]{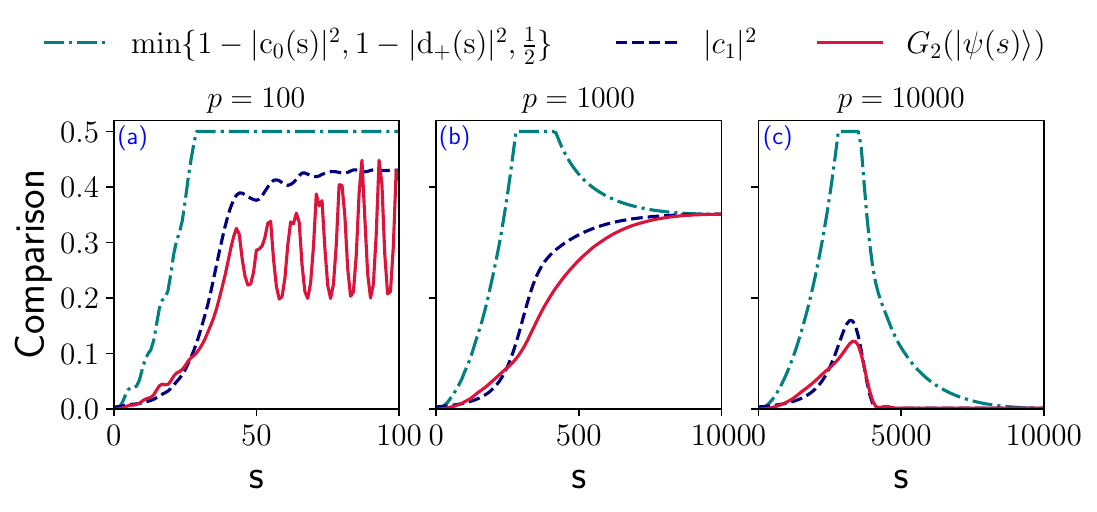}
    \caption{The GGM $G_2$ (solid red) is upper bounded by $\min{\left\{1 - |c_0(s)|^2, 1- |d_+ (s)|^2,\frac{1}{2}\right\}}$ (dashdotted teal). When the sweep speed is slow, i.e., for $p=1000,10000$, the instantaneous state is a superposition of a few states toward the end of the anneal, and $G_2$ and $|c_1|^2$ (blue dashed) thus become almost equal, as seen in the panels (b,c).}
    \label{fig:overlap and GGM}
\end{figure}

In Fig.~\ref{fig: fixed_instance}(a,b), the GGM and the occupation probability of the first excited state $|c_1|^2$ are quite comparable throughout the sweep, especially for larger $p$.
This similarity becomes even more apparent by plotting $G_2$ and $|c_1|^2$ together in Fig.~\ref{fig:overlap and GGM}. 
This quantitative comparison shows that the GGM and $|c_1|^2$ do neither strictly coincide with each other nor does one strictly bound the other.  
Rather, $G_2$ is strictly bounded from above by the overlap probability with the initial state and the exact solution state, as we now prove.

Calculating the GGM of a given state $\ket{\psi}$ is equivalent to finding the two-party product state $\ket{\pi}_{\mathcal{A}:\mathcal{B}}$ with minimal distance from $\ket{\psi}$, see Eq.~\eqref{eq: GGM_using_schmidt}. 
As the annealing sweep terminates with the cost Hamiltonian, whose eigenstates $\ket{e_i}$, defined in Eq.~\eqref{eq: expansion in eigenstate} are product states, it seems reasonable to estimate the GGM by using these as a basis. 
The true GGM has to be smaller than the distance to any of these, i.e., 
\begin{equation}
   G_2\left(\ket{\psi(s)}\right) 
   \leq 1- \underbrace{|\langle e_i | \psi(s)\rangle|^2}_{|c_i(s)|^2} , \quad \forall i\,.
\end{equation}
Towards the end of a slow sweep, the expansion of Eq.~\eqref{eq: expansion in eigenstate} will have the largest weight in the ground-state of the cost Hamiltonian $\ket{e_0}$. We thus estimate the GGM by the upper bound 
\begin{equation}\label{eq:G_2 and c_0}
    G_2\left(\ket{\psi(s)}\right) \leq 1-|c_0(s)|^2.
\end{equation}

Similarly, during the early stages of the sweep, the wave function will have a large weight in the initial product state $\ket{+}^{\otimes N}$. 
It is thus reasonable to expect the distance to this state to provide a good upper bound in the initial part of the anneal, 
\begin{equation}
    G_2 \left(\ket{\psi(s)}\right)  \leq 1- |(\bra{+}^{\otimes N})  |\psi(s)\rangle|^2 \equiv 1- |d_+ (s)|^2\,.
\end{equation}

Moreover, the GGM for qubits has a theoretical maximum of $1/2$~\cite{footnoteMax}. 
Putting all three bounds together, the GGM is upper bounded by 
\begin{equation}
    \label{eq:fullupperbound}
    G_2 \left(\ket{\psi(s)}\right)  \leq \min{\left\{1 - |c_0(s)|^2, 1- |d_+ (s)|^2,\frac{1}{2}\right\}}.
\end{equation}

Numerical experiments exemplify this bound, see Fig.~\ref{fig:overlap and GGM}. 
Considering the simplicity of the bound, the qualitative agreement is very satisfactory. In particular, the bound is saturated by $1-|c_0|^2$ towards the end of a successful anneal, as $|c_0|^2$ becomes the largest probability among all $|c_i|^2$s. Later, in Sec.~\ref{sec: final GGM and success probability}, we further use Eq.~\eqref{eq:G_2 and c_0}, to relate the success probability, i.e., the final $|c_0|^2$, to the final GGM. 
Importantly, the values entering Eq.~\eqref{eq:fullupperbound} are easily obtained experimentally by projecting the state onto the computational basis or a locally rotated basis. It can thus be used as a physically-informed proxy for bounding multipartite entanglement.

As remarked above, in addition there seems to be a qualitative similarity between the first excited state overlap probability $|c_1(s)|^2$ and the GGM $G_2(\psi)$, see Fig.~\ref{fig:overlap and GGM}. Towards the very end of the sweep for $p=1000, 10000$, this similarity can be explained as only the states $\ket{e_0}$ and $\ket{e_1}$ remain prominent, i.e., $|c_0|^2 +|c_1|^2 \simeq 1$, and the upper bound of the GGM saturates, thus giving $G_2=1-|c_0|^2 \simeq |c_1|^2$. However, their similarity in the middle of the sweep cannot be explained solely by this reasoning. Nonetheless, we speculate that such non-zero $|c_1|^2$ has a role in keeping the instantaneous state entangled, as---assuming $|c_1|^2<|c_0|^2$---it heralds the presence of an entangled superposition state.

\section{Genuine multipartite entanglement barrier in Random Max-Cut ensemble}\label{sec: entanglement barrier}
\begin{figure*}[hbt!]
    \centering
    \includegraphics[width=1\textwidth]{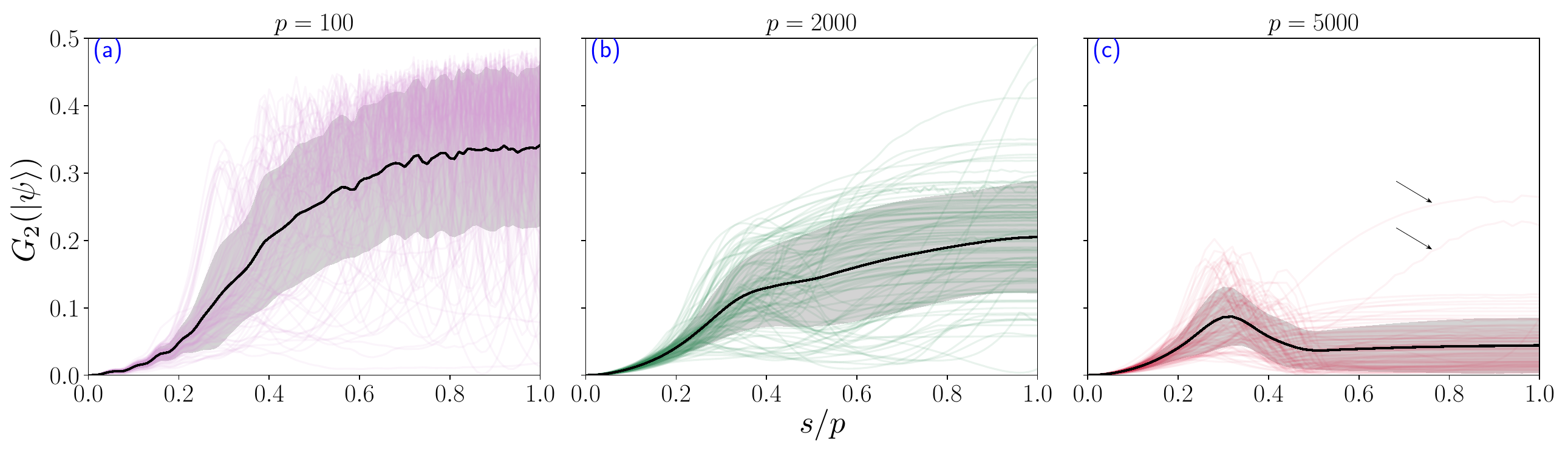}
    \caption{GGM from numerical benchmarks of 100 MaxCut instances. Panels (a) to (c), correspond to TQA with $p=100,2000,5000$ layers, respectively, i.e., increasing annealing time. For $p=100, 2000$, although some individual instances have peaks in the GGM (refer to Fig.~\ref{fig: fixed_instance} for an example), the average GGM does not show any peak (panel a) or only a small hump (b). In contrast, for $p=5000$ (c), a peak in the GGM is visible also on average. With increasing $p$, the fluctuations in GGM between different instances decrease. Problem instances with no barrier are indicated by arrows. Again, strong oscillations typical for the defects generated by fast sweeps appear in $p=100$.
    }
    \label{fig:GGM for different instances}
\end{figure*}

In Sec.~\ref{sec: fixed problem}, we have observed that the GGM assumes a peak during a slow quantum-annealing sweep, see Fig.~\ref{fig: fixed_instance}(d).
We call this a multipartite entanglement barrier since the inability to build up and then reduce the GGM prevents the algorithm from preparing the non-entangled solution state. 
We now investigate whether the observed barrier is (i) a universal feature not specific to the MaxCut problem instance used above 
and (ii) whether it is related to the instantaneous energy gap.

We analyze 100 random MaxCut instances for 8 vertices of 6-regular graphs with edge weights sampled from a Gaussian distribution $\mathcal{N}(\mu=6, \sigma=3)$. 
We run TQA with annealing sweep speeds $p=100, 2000, 5000$. 
For $p = 100$, many instances display a peak or an oscillation in the GGM.
However, there is no peak on average since the peaks occur at different times for different instances, see Fig.~\ref{fig:GGM for different instances}(a). 
Notably, the spread in the GGM at the beginning of the protocol is small.
This trend is also observed for $p = 2000$ and $p = 5000$.

For $p = 2000$, we observe a small hump also in the instance-averaged GGM around $s/p=1/3$, see Fig.~\ref{fig:GGM for different instances}(b).
The final state's GGM is lower as compared to $p = 100$. 
The entanglement peak, or barrier, becomes more strongly apparent at $p = 5000$, see Fig.~\ref{fig:GGM for different instances}(c).  
In this case, the multipartite entanglement remains relatively low throughout the sweep, and the final GGM is minimal, indicating that the protocol has reached the solution state with high probability. 
Nevertheless, there are problem instances for which the GGM increases until $s/p=1$, e.g., see the two red lines in Fig.~\ref{fig:GGM for different instances}(c), which do not exhibit a multipartite entanglement barrier. 

For comparison, we also investigate the GGM in a QAOA protocol with few layers but classically optimized angles $\beta$ and $\gamma$ in Eq.~\eqref{eq:Hstep}, see App.~\ref{app: GGM in QAOA} for details.
Here, we do not observe a peak in the GGM but rather a build-up of the GGM with each additional QAOA layer.
This behavior is similar to a short-depth quantum anneal, as in Fig.~\ref{fig:GGM for different instances}(a), which is reasonable considering the few layers used in QAOA.

\subsection{Relation between GGM barrier and energy gap}

In equilibrium, high entanglement can appear at quantum critical points with small energy gaps~\cite{orus2004universality, hauke2016measuring}. 
However, this connection between low-lying spectrum and entanglement may not necessarily be expected in the out-of-equilibrium context of quantum optimization.
Here, we investigate (a) whether the point where the energy gap closes correlates with the position of the GGM peak and (b) whether the size of the energy gap relates to the amount of GGM generated.

\begin{figure}[hbt!]
    \centering
    \includegraphics[width=1\columnwidth]{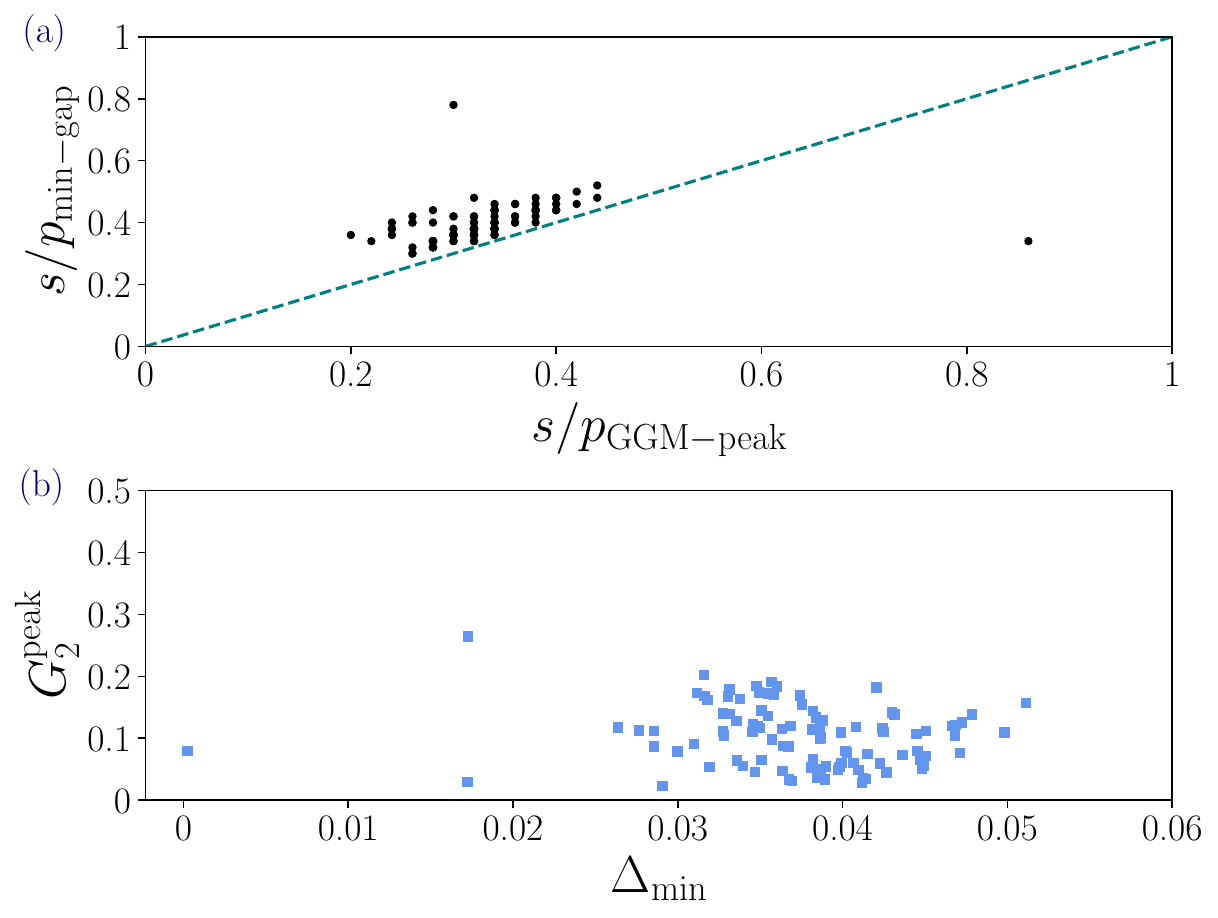}
    \caption{Relation between the multipartite entanglement barrier and the energy-gap 
    for 100 instances solved with TQA with $p=5000$ layers and $dt=0.2$. 
    (a) The time of the smallest energy gap in the sweep versus the time at which the GGM peaks are strongly correlated (at the 99\% confidence level). 
    (b) Such strong correlations are not observed for the magnitude of the GGM peak versus the size of the minimal energy gap (the data yields a weak anti-correlation at the 80\% confidence level). 
    }
    \label{fig:gap and GGM}
\end{figure}

In Fig.~\ref{fig:gap and GGM}(a), we plot the time $s/p$ at which the energy gap and the GGM are minimum and maximum, respectively, for the same 100 problem instances as in Fig.~\ref{fig:GGM for different instances}. 
We report the first peak in the GGM to avoid ambiguities in case the GGM oscillates.
The position of the minimum gap has a linear correlation with the position of the first GGM peak. 
Quantitatively, Pearson's correlation coefficient is $r=0.28$  
for the full dataset, including the two extreme points. 
The corresponding $t$-score is $t=|r|\sqrt{(n-2)/ (1-r^2)}=2.89$, with $n=100$ problem instances. 
This implies a statistically significant correlation at the 99\% confidence level.
Excluding the two outliers increases Pearson's correlation coefficient to $r=0.525$. 
By contrast, we do not observe such a strong correlation between the magnitude of the energy gap and the size of the GGM peak, see Fig.~\ref{fig:gap and GGM}(b).
Indeed, the associated correlation coefficient of $r=-0.16$ is only significant at the 80\% confidence level. 
As in this out-of-equilibrium situation the near-degeneracy between ground and first excited state is not the sole reason for entanglement, this absence of a strong correlation between the amount of entanglement and the energy gap is not unexpected.

\subsection{Relation to Success Probability}

Generating and removing entanglement in quantum circuits are complex operations. 
For our MaxCut problems with a non-degenerate solution, quantum annealing must eliminate any existing entanglement to achieve the final product state. This raises an interesting question: does the non-zero entanglement generated during quantum annealing hinder the optimization algorithm's ability to find the optimal solution? Specifically, does a higher GGM in the final state limit the probability of success? If so, given the presence of an entanglement barrier, could a drop of the GGM from its peak enhance the chances of achieving a better success probability?

\subsubsection{Final GGM and success probability}\label{sec: final GGM and success probability}

\begin{figure}[h]
    \centering
    \includegraphics[width=1\columnwidth]{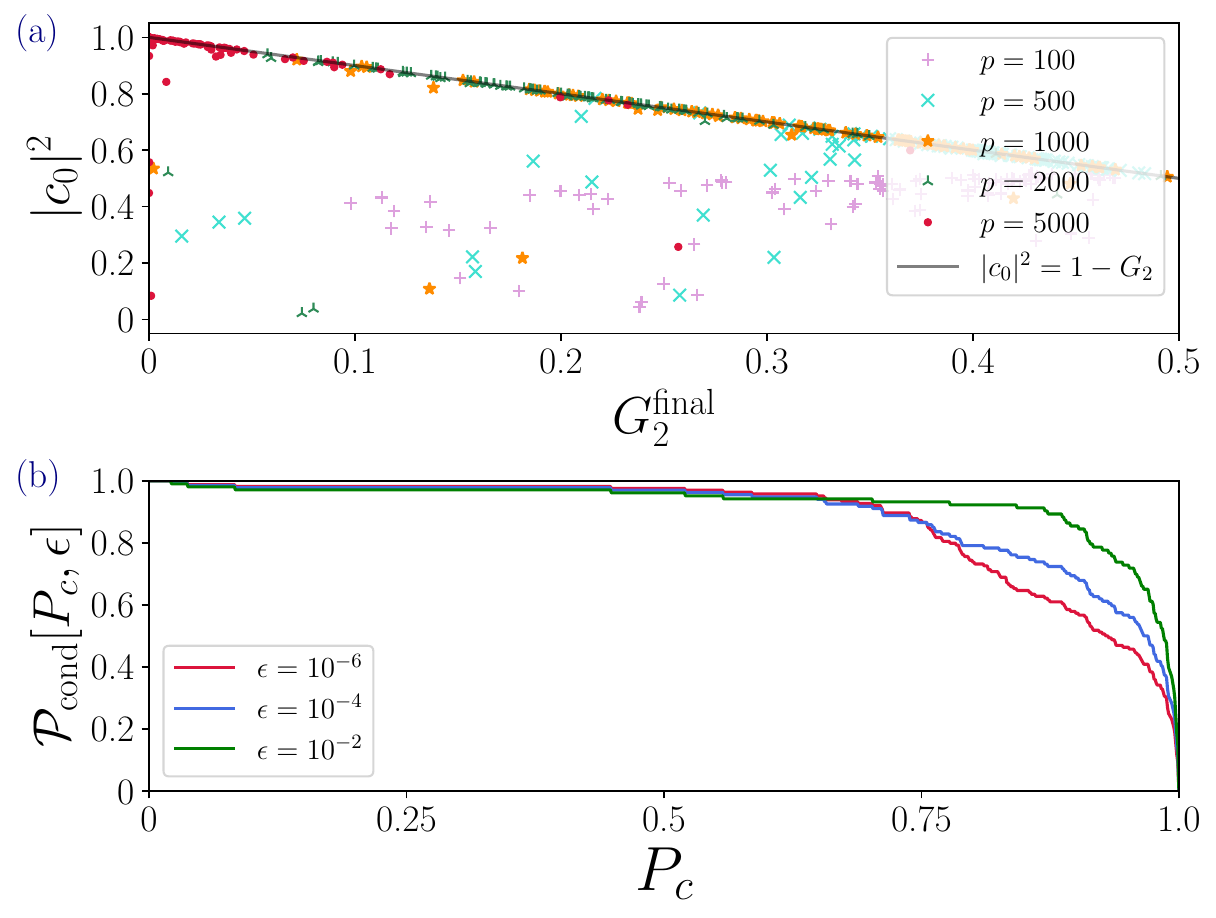}
    \caption{Relation of entanglement and success probability, over 100 instances of 6-regular graphs with 8 vertices. (a) The final success probability $|c_0|^2$ is bounded from above by $1-G_2^{\mathrm{final}}$. 
    Large $p$ saturate the bound. 
    (b) The conditional probability of obtaining a success probability $>P_c $ increases if the amount of disentanglement ($\Delta G$) increases, exemplified with $\epsilon > \{10^{-6}, 10^{-4}, 10^{-2} \}$ for 100 instances each with ($p=2000, 5000$) } 
    \label{fig:success probability and ggm} 
\end{figure}

As a corollary to the Sec.~\ref{sec: excited state and GGM}, the success probability, i.e., the probability corresponding to the ground-state of the cost Hamiltonian, $|c_0|^2$, is bounded above by $1-G_2^{\rm final}$. In Fig.~\ref{fig:success probability and ggm}(a), we see that with an increasing number of layers (slower sweep), the success probability saturates the bound. In contrast, with fewer layers, the success probability does not saturate the bound, as the final state has support on multiple eigenstates, making the bound of Eq.~\eqref{eq:G_2 and c_0} using a single amplitude too loose. These results are similar to the observed relation between \textit{bipartite} entanglement for a fixed bipartition and the success probability in Ref.~\cite{hauke2015probing}.

\subsubsection{Disentanglement and success probability}
As a complementary analysis, we also study how the amount of dis-entanglement, i.e., the reduction of GGM after its peak, $\Delta G= G_2^{\rm max}- G_2^{\rm final}$, is related to the success probability. 
To this end, we calculate the conditional probability ($\mathcal{P}_{\rm cond}[P_c, \epsilon]$) of obtaining a success probability greater than some $P_c$, given that the sweep has shown some minimum amount of disentanglement, $\Delta G > \epsilon$
\begin{equation}
    \mathcal{P}_{\rm cond}\Big[|c_0|^2 > P_c \big| \Delta G > \epsilon\Big]= \frac{\mathcal{P}\Big[(|c_0|^2 > P_c) \cap (\Delta G > \epsilon)\Big]}{ \mathcal{P} \Big[ \Delta G > \epsilon\Big]}.
\end{equation}

We study only the cases where a GGM barrier is observed, i.e., $G_2^{\rm max} > G_2^{\rm final}$.  
We choose $p = 2000, 5000$, computing the TQA of 100 instances for both numbers of layers.
To show reasonable comparison, we need to choose small values of $\epsilon=(10^{-6}, 10^{-4}, 10^{-2})$, as the final GGM is itself small in terms of absolute values. 
We find a larger $\Delta G$ (increasing $\epsilon$) to result in a larger success probability, see Fig.~\ref{fig:success probability and ggm}(b). 
In particular, large values of $P_c$ are capped by the given fixed minimum disentanglement, leading to a downward bend in the curves [Fig.~\ref{fig:success probability and ggm}(b)] and suggesting the necessity of a large amount of disentanglement for a good success probability.

\section{Conclusion}\label{sec: conclusion}
To summarize, multipartite entanglement arises in Trotterized quantum annealing even when the solution of the classical optimization problem is non-degenerate. While fewer layers---or a faster sweep---generate more entanglement, increasing the number of layers causes the GGM to grow, reach a peak, and then decline to zero during the sweep---a phenomenon one can call a multipartite entanglement barrier. 
We prove that the success probability is upper-bounded by $1-G^{\rm final}_2$, which is saturated when the final state is close to a product state (such as the desired solution state), as illustrated by numerical results from MaxCut instances. 
Thus, achieving a higher success probability typically requires lower final-state entanglement. 
Such a relation can be leveraged to benchmark quantum hardware~\cite{miessen2024benchmarking, santra2024squeezing, azar2024calibrating, teramoto2023role}.
From the phenomena of the entanglement barrier, we quantify the relation between disentanglement and success probability through a conditional probability and find that a stronger drop in multipartite entanglement after the GGM peak improves the success probability. 
Moreover, as the GGM corresponds to an optimal bipartition with minimal bipartite entanglement~\cite{buerschaper2014topological}, insights about GGM can be used in circuit-cutting protocols to find an optimal cut such that the multipartite entanglement loss is minimal~\cite{hart2024quantum}.

Our work has consequences for short-depth protocols such as QAOA and its variants.
The classical parameter optimization loop is time-consuming~\cite{Weidenfeller2022}, and navigating the optimization landscape in variational quantum algorithms is NP-hard~\cite{bittel2021nphard}.
Therefore, recent research investigates methods to obtain QAOA parameters by classical means.
In particular, tensor networks may help to obtain good parameters~\cite{Streif2020}.
Here, it is sufficient to qualitatively approximate the loss landscape to produce good QAOA parameters.
Our work shows that multipartite entanglement can build up in QAOA.
An interesting research question is thus whether tensor-network methods can provide a qualitative approximation of the loss landscape that is good enough to produce good QAOA parameters at a classically tractable bond dimension.

Further research could explore how the multipartite entanglement barrier manifests if the Hamiltonian is engineered using the tools from shortcuts to adiabaticity~\cite{guery2019shortcuts}. 
By exploring possible links to the graph's spectral properties, one may hope to understand better why GGM barriers cluster in specific regions~\cite{broder1987second}. Moreover, understanding the source of such entanglement in terms of physical phenomena, e.g., the Landau--Zener transition~\cite{ivakhnenko2023nonadiabatic} or multiqubit quantum tunneling~\cite{boixo2016computational} one may hope to predict better the possibility or impossibility of reaching the solution state~\cite{altshuler2010anderson}.  In the future, it would also be useful to investigate the roles of other quantum resources in quantum optimization, such as non-stabilizerness~\cite{veitch2014resource}, discord~\cite{Bera_2017}, and coherence~\cite{Streltsov_2017}.
Our work focuses on pure states. 
However, some quantum algorithms may leverage mixed states ~\cite{knill1998power,aharonov1998quantum,datta2007role,schulte2012control}, and it will be interesting to use the GGM to study the role of multipartite entanglement in such cases~\cite{das2016generalized,wei2003geometric,vollbrecht2001entanglement}.

\section{Acknowledgment}
This project has received funding by the European Union under Horizon Europe Programme - Grant Agreement 101080086 - NeQST and under NextGenerationEU via the ICSC – Centro Nazionale di Ricerca in HPC, Big Data and Quantum Computing. This project has received funding from the Italian Ministry of University and Research (MUR) through the FARE grant for the project DAVNE (Grant R20PEX7Y3A), was supported by the Provincia Autonoma di Trento, and by Q@TN, the joint lab between the University of Trento, FBK-Fondazione Bruno Kessler, INFN-National Institute for Nuclear Physics and CNR-National Research Council. 
D.J.E.\ acknowledges funding within the HPQC project by the Austrian Research Promotion Agency (FFG, project number 897481) supported by the European Union – NextGenerationEU.
Views and opinions expressed are however those of the author(s) only and do not necessarily reflect those of the European Union or the European Commission. Neither the European Union nor the granting authority can be held responsible for them. S.S.R. acknowledges the faculty research scheme at IIT (ISM) Dhanbad, India under Project No. FRS/2024/PHYSICS/MISC0110.

\section{Data Availability}
The data presented in this article is available from Zenodo~\cite{santra_zenodo}.

\appendix 
\section{Details of the graphs used in main text}\label{app: problem instance}
\begin{figure}[hbt!]
    \centering
    \includegraphics[width=0.7\columnwidth]{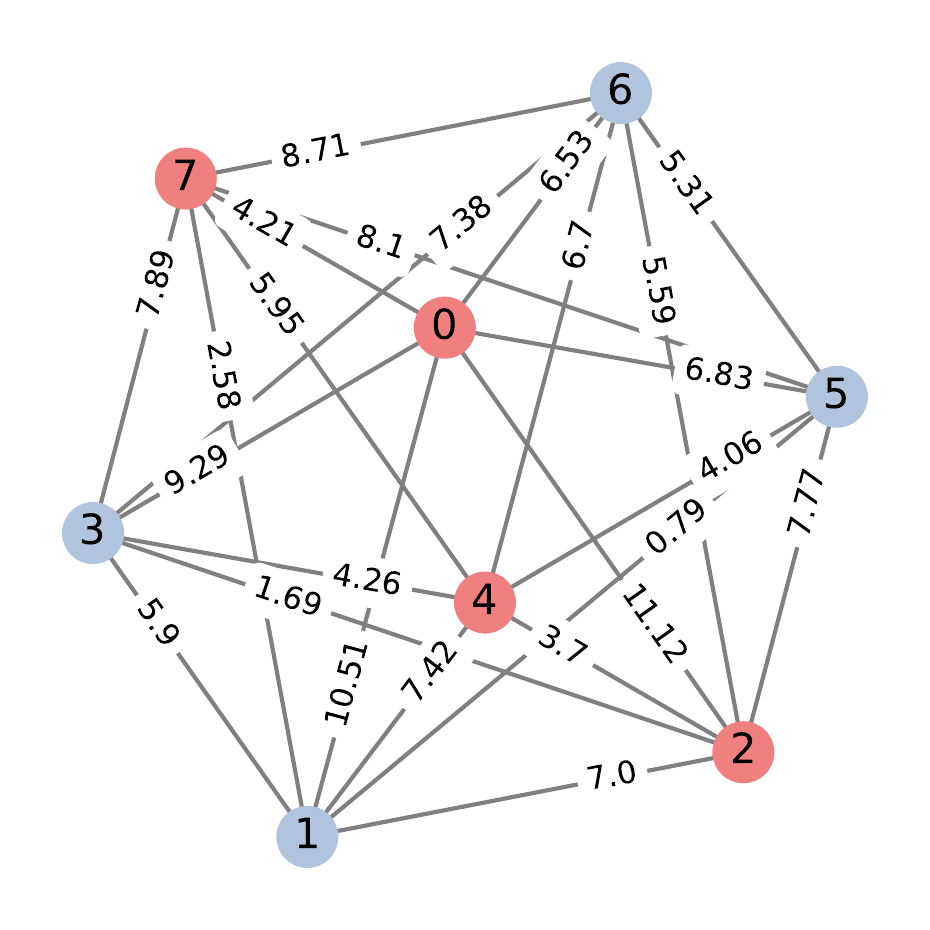}
    \caption{
    Graph instance used in Sec.~\ref{sec: fixed problem}. Weights $J_{ij}$ are indicated on the edges.
    The MaxCut solution is denoted by red and blue node colors.}
    \label{fig: graph}
\end{figure}

\begin{figure}[hbt!]
    \centering
    \includegraphics[width=0.9\columnwidth, clip, trim=0 400 35 0]{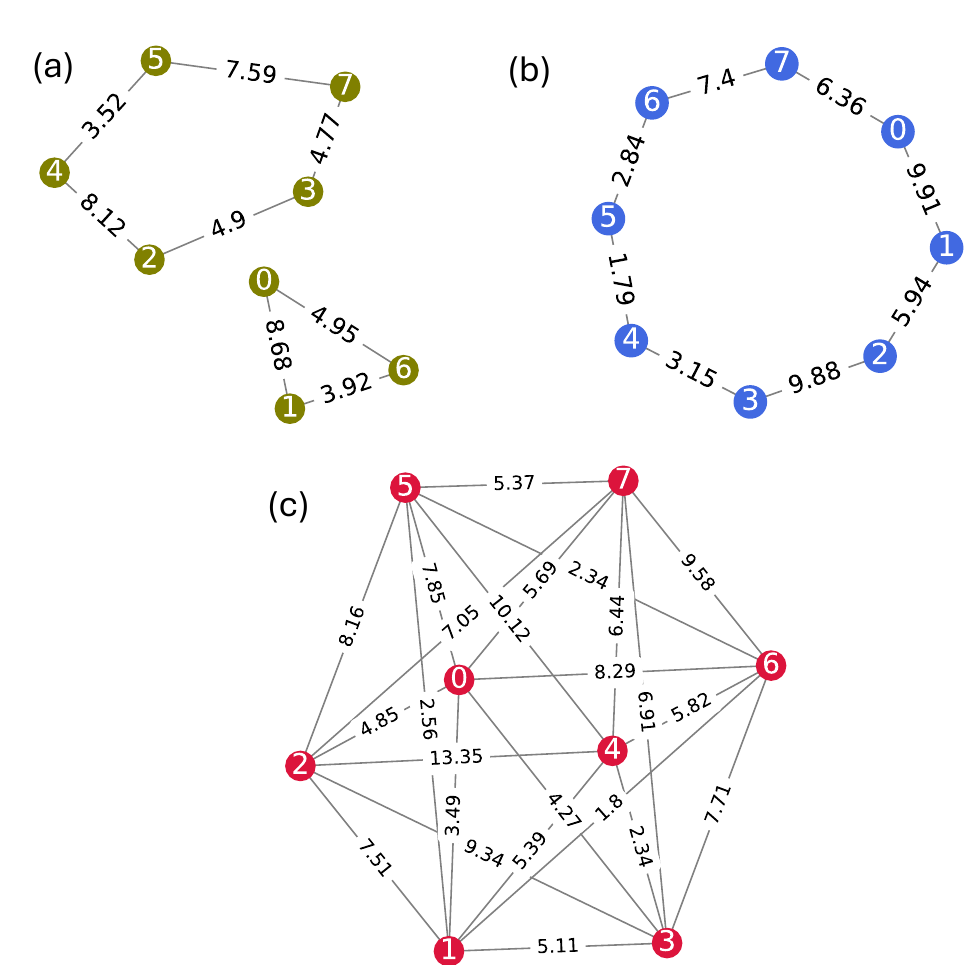}
    \caption{Edge-weights of the graphs used in Fig.~\ref{fig: GGM vs VNE} of Sec.~\ref{sec: GGM in fixed problem}: (a) two-regular disconnected graph with two rings, (b) two-regular connected graph, and (c) six-regular connected graph.}
    \label{fig:3_graph for comparision}
\end{figure}

Here, we specify the six-regular eight-vertex graphs used in the simulations in Sec.~\ref{sec: fixed problem}.
All edge weights $J_{ij}$ are chosen from a Gaussian distribution $\mathcal{N}(6,3)$. 
Figure~\ref{fig: graph} shows the resulting six-regular eight-node graph instance used in Fig.~\ref{fig: fixed_instance}. 
The solution of this MaxCut instance is defined by the separation into vertex sets $(1,3,5,6)$ and $(0,2,4,7)$.
The graphs underlying the data in Fig.~\ref{fig: GGM vs VNE}, which show the difference between the von Neumann entanglement entropy and the GGM, are drawn in Fig.~\ref{fig:3_graph for comparision}.

\section{$\Delta t$ for TQA}\label{app: best dt}
\begin{figure}
    \centering
    \includegraphics[width=1\columnwidth]{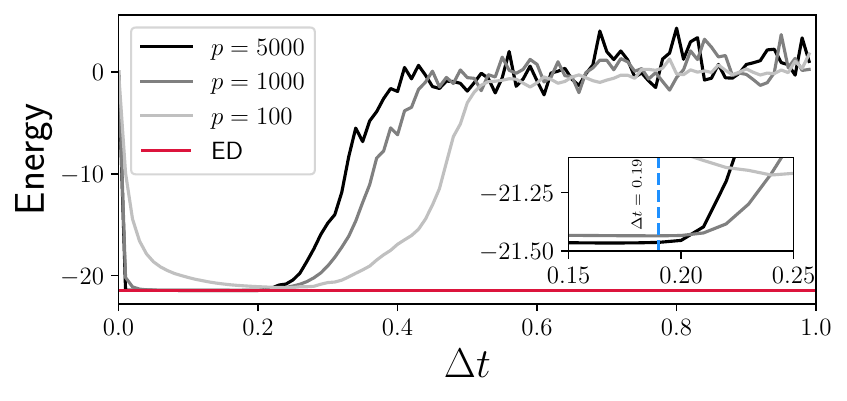}
    \caption{Energy of the final state of Trotterized quantum annealing as a function of $\Delta t$ for different Trotter layers $p$ for the MaxCut problem of Fig.~\ref{fig: graph}, compared to the ground-state energy obtained from exact diagonalization (ED).}
    \label{fig: best dt}
\end{figure}

In Trotterized quantum annealing (TQA), the quality of the annealing, which is related to the success probability, depends on the total number of Trotter layers $p$ and the chosen time step $\Delta t$, see Eq.~\eqref{eq: delta t}. 
In principle, $\Delta t$ should be chosen to maximize the success probability. 
However, to study the impact of the sweep time (i.e., the number of layers multiplied by the step size $\Delta t$) on the behavior of the GGM, we choose a value of $\Delta t$ that keeps the energy of the final state small for all considered numbers of layers ($p=100, 1000, 5000$). 
Our simulations show that there is no $\Delta t$ value that simultaneously minimizes the energy of all layers $p$, see Fig.~\ref{fig: best dt}.
As a compromise, we choose $\Delta t=0.19$ for Fig.~\ref{fig: fixed_instance}, for which the final energy is close to the absolute ground state for all $p$ considered. 
In Fig.~\ref{fig:GGM for different instances}, when we average over 100 instances, we use $\Delta t=0.2$ for all instances instead of optimizing $\Delta t$ for each instance. 

\section{Degeneracy breaking and pairing of states}\label{app: breaking degeneracy}
To break the $\mathbb{Z}_2$ symmetry in MaxCut, we add an offset term as described in Eq.~\eqref{eq: max_cut with symmetry breaking}. 
Here, we show how the degeneracy in the spectrum of $H_\mathrm{C}$ is lifted with increasing offset. We choose $0.05$ as offset to lift the degeneracy while preserving the eigenlevel structure.

\begin{figure}[h!]
    \centering
    \includegraphics[width=\columnwidth]{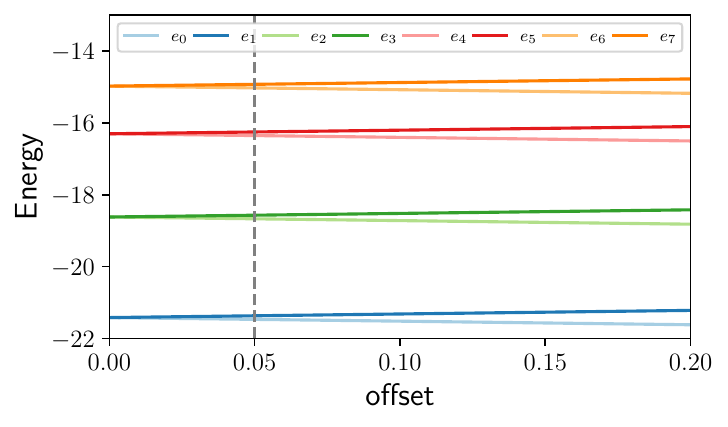}
    \caption{A non-zero magnitude $f$ of the offset $fZ_0$ lifts the degeneracy in the eigenvalue of $H_\mathrm{C}$, as illustrated for the instance described in Fig.~\ref{fig: graph}. As $f$ increases, the paired states become more and more distinguishable. For the data shown in the main text, we choose an offset of $f=0.05$ (grey vertical line) at which the states are non-degenerate but still form pairs close to each other in energy.}
    \label{fig:offset_effect}
\end{figure}

\begin{figure}
    \centering
    \includegraphics[width=1\columnwidth]{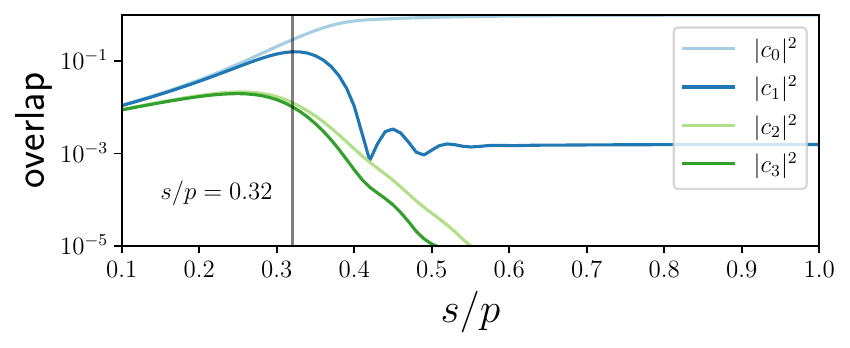}
    \caption{Population of the four lowest-energy instantaneous eigenstates, for the MaxCut problem from Fig.~\ref{fig: graph} and with $p=5000$. Pairing is observed until the barrier, after which the pairs break, and $|c_0(s)|^2$ approaches $1$.}
    \label{fig: states coming in pair}
\end{figure}

At this level of bias, although the $\mathbb{Z}_2$-symmetric states are no longer degenerate, they are still energetically close and thus have a pairing effect during the annealing sweep. For the graph instance described in Fig.~\ref{fig: graph}, we study the overlap probability of the instantaneous state with the four lowest energy states $\ket{e_i}$, $i\in\{0,1,2,3\}$ of $H_\mathrm{C}$ during the sweep.
Until the sweep reaches the GGM peak, shown in Fig.~\ref{fig: fixed_instance}(a), we see that the population of the ground and the first excited states are almost equal, see Fig.~\ref{fig: states coming in pair}.
The same holds for the third and fourth excited states.
This hints at a pairing of $\ket{e_0}$ with $\ket{e_1}$ and of $\ket{e_2}$ with $\ket{e_3}$.
Moreover, again due to the $\mathbb{Z}_2$ symmetry, $|e_0\rangle$ and $|e_1\rangle$ are bit-wise orthogonal basis states (and the same for $\ket{e_2}$ and $\ket{e_3}$).
An equal superposition of such states creates a GHZ-like state with a high GGM. 
For example, with the graph considered here $|e_0\rangle = |10101001\rangle$ and $|e_1\rangle=|01010110\rangle$ form the state $|\mathrm{GHZ}_8\rangle=\frac{1}{\sqrt{2}} (|e_0\rangle + |e_1\rangle)$ state. 

\section{Optimal bipartitions corresponding to the GGM}\label{app: which bipartition}
 
\begin{figure*}
    \centering
    \includegraphics[width=0.75\textwidth, clip, trim=0 230 0 0]{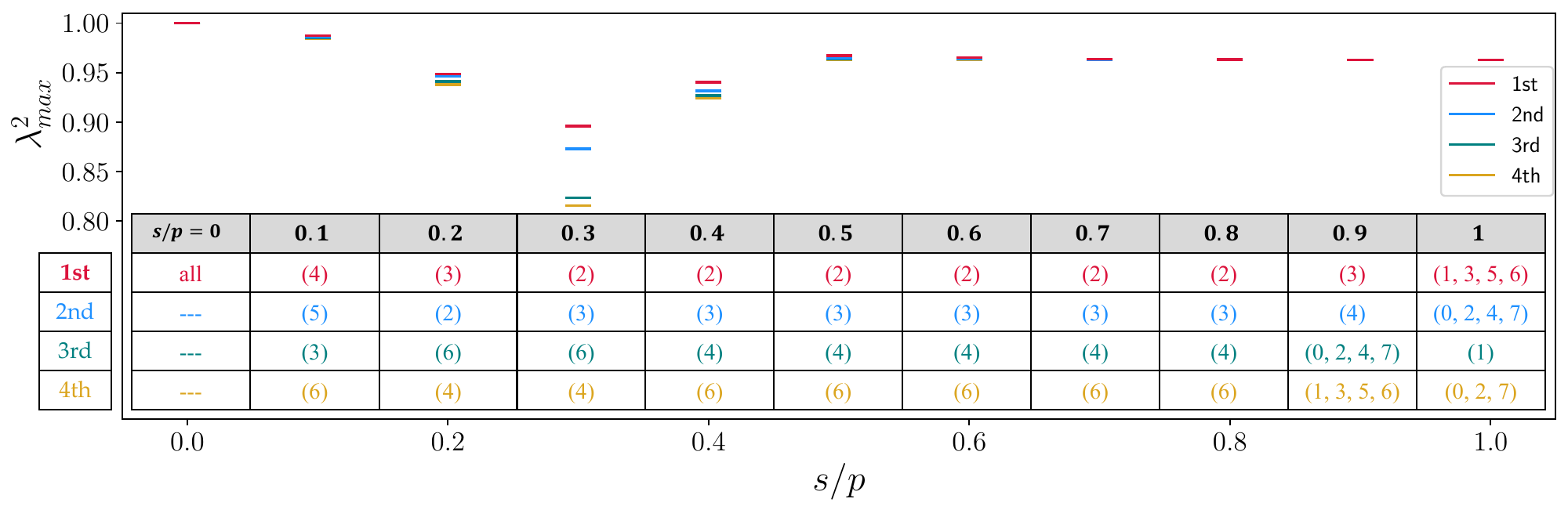}
    \caption{ Largest Schmidt coefficients along the annealing sweep, at selected times for the partitions $(\mathcal{A}):({\mathcal{B}})$ yielding the four largest values, for the MaxCut problem of Fig.~\ref{fig: graph}. The table gives the corresponding sets of qubits in $\mathcal{A}$. The maximum Schmidt values decrease from the initial states until $s/p=0.3$, implying an increasing GGM. Later, the Schmidt values increase while the optimal bipartitions remain unchanged until $s/p=0.8$. Subsequently, the GGM remains constant while the optimal bipartition changes to finally reach $\mathcal{A}=(1,3,5,6)$.}

    \label{fig:which bipartitions}
\end{figure*}

To find the value of the GGM, we need to look for the bipartition that yields the maximum Schmidt coefficient. A question that may naturally arise is whether such a bipartition is unique or changes with the annealing sweep.
At the start of the anneal, at $s/p=0$, we have $\lambda^2_{max}=1$ in all bipartitions. This implies a vanishing GGM, see Fig.~\ref{fig:which bipartitions}. As the annealing schedule proceeds, 
$\lambda^2_{max}$ of different partitions starts to differ; for example, at $s/p=0.2,$ partition $\mathcal{A}=(3), {\mathcal{B}}=(0,1,2,4,5,6,7)$ and partition $\mathcal{A}=(2), {\mathcal{B}}=(0,1,3,4,5,6,7)$ have different values of $\lambda^2_{max}$. The individual Schmidt values decrease during the sweep until they reach the lowest $\lambda^2_{max}$ corresponding to the GGM barrier at around $s/p=0.3$. 
After this point, the GGM decreases, and the partition producing the $\lambda^2_{max}$ --- $\mathcal{A}=(2), {\mathcal{B}}=(0,1,3,4,5,6,7)$ --- does not change until $s/p=0.8$. 
Towards the end of the anneal, in the range $s/p>0.8$, the maximum Schmidt values and the GGM do not change considerably. However, the optimal bipartition varies until it finally reaches $\mathcal{A}: {\mathcal{B}}=(1,3,5,6):(0,2,4,7)$. 
This bipartition has the maximum Schmidt values (maximal bipartite entanglement). In summary, it is interesting to see that from $s/p=0.1$ to $0.3$, the GGM changes, but the corresponding partition does not. Thereafter, and until the end of the anneal, the GGM remains constant, but the underlying partition changes.

\section{GGM in Quantum Approximate Optimization Algorithm}\label{app: GGM in QAOA}

In this work, we are mostly interested in Trotterized quantum annealing. Nonetheless, it is interesting to compare the results with those obtained with a QAOA protocol. 
The standard QAOA applies $p$ layers of the unitaries $\exp(-i\beta_k H_\mathrm{M} )\exp(-i\gamma_k H_\mathrm{C})$, with $k = 1,..., p$ on the initial state $\ket{+}^{\otimes N}$ to create a trial state $\ket{\psi(\boldsymbol{\gamma}, \boldsymbol{\beta})}$. 
Then, a classical optimizer seeks the optimal parameters $\boldsymbol\beta=(\beta_1,\dots,\beta_p)$ and $\boldsymbol\gamma=(\gamma_1,\dots,\gamma_p)$ that minimize the energy $\braket{\psi(\boldsymbol\beta,\boldsymbol\gamma)|\hat{H}_C|\psi(\boldsymbol\beta,\boldsymbol\gamma)}$, which is measured by the quantum processor. 
As optimizing over $2p$ parameters in the classical optimizer is a bottleneck, we limit our simulation to small depths $p$. 

In its spirit, QAOA is an approximate algorithm. It is designed to provide samples of a good set of solutions using only short-depth circuits (in contrast to quantum annealing, where one is typically more interested in slow but near-optimal runs). 
Therefore, as a figure of merit to quantify the performance of QAOA, we use the approximate ratio $\alpha=\langle E \rangle/ E_{\rm exact}$ instead of success probability. 
We run QAOA with different depths $p$ for the same MaxCut instance we consider in Sec.~\ref{sec: fixed problem}, see App.~\ref{app: problem instance}. 

For fixed depth $p$, the GGM increases with each QAOA layer $k$; however, for larger $p$, there is a kink at $k=3$. 
Nonetheless, the approximation ratio $\alpha$ and the final GGM increase with QAOA depth $p$ without a clear presence of an entanglement barrier. 
This is akin to the observed feature in the GGM when the number of TQA layers is small, see $p=100$ in Fig.~\ref{fig:GGM for different instances}(a) of the main text.
Recall that the multipartite entanglement barrier appears when the instantaneous overlap probability of the ground state $|c_0|^2$ and the first excited state $|c_1|^2$ start to become distinguishable, see Fig.~\ref{fig: fixed_instance}(b) and Fig.~\ref{fig: states coming in pair}. 
Thus, a large GGM may not necessarily hinder its performance as long as it stems from the superposition of the true ground state $\ket{e_0}$ with a few excited states ${\ket{e_{i>0}}}$. In that case, 
there is a high probability that a manageable number of repetitions of the algorithm will then permit us to measure the optimal state on the computational basis at least once.

\begin{figure}[]
    \centering
    \includegraphics[width=\columnwidth]{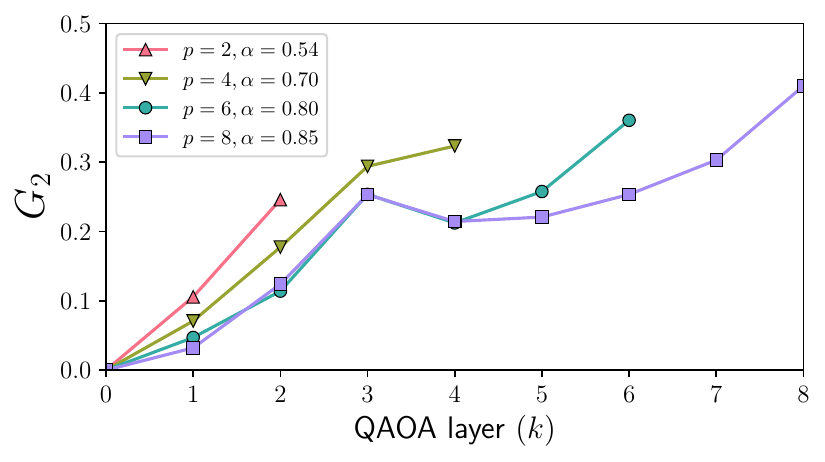}
    \caption{The GGM ($G_2$) of the state obtained after layer $k=0,\dots,p$ of a depth-$p$ QAOA. The approximation ratio (see legend) and final GGM both increase with $p$.}
    \label{fig:GGM_QAOA}
\end{figure}

\newpage


\begin{thebibliography}{69}
\bibitem [{foo()}]{footnoteMax}%
  \BibitemOpen
  \href@noop {} {}\bibinfo {note} {We can understand this intuitively as follows. Consider an arbitrary bipartition \(\mathcal{A}:\mathcal{B}\) where \(|\mathcal{A}| = m\) and \(|\mathcal{B}| = N - m\). For this bipartition, there can be at most \(d^m\) non-zero Schmidt coefficients, denoted as \(\{\lambda^1_{\mathcal{A}:\mathcal{B}}, \lambda^2_{\mathcal{A}:\mathcal{B}}, \dots, \lambda^{k=d^m}_{\mathcal{A}:\mathcal{B}}\}\). To maximize the GGM, we need to arrange these Schmidt coefficients so that \(1 - (\lambda^1_{\mathcal{A}:\mathcal{B}})^2\) is as large as possible. This is achieved when all \(\lambda^k_{\mathcal{A}:\mathcal{B}}\) are equal, i.e., \(\lambda^k_{\mathcal{A}:\mathcal{B}} = \frac{1}{\sqrt{d^m}}\) for all \(k\). Therefore, among all possible bipartitions, the maximum of \(1 - (\lambda^1_{\mathcal{A}:\mathcal{B}})^2\) is obtained in the case where \(m = 1\), and the maximally possible GGM value is \(G_2 = \frac{1}{d}\).}\BibitemShut {Stop}%
\end{thebibliography}
\end{document}